\documentclass[prl,twocolumn,showpacs,noshowkeys,superscriptaddress,preprintnumbers,amsmath,amssymb]{revtex4}

\usepackage{graphicx}
\usepackage{dcolumn}
\usepackage{amsmath}
\usepackage{eucal}

\begin{document}

\title{Spin Squeezing via One-Axis Twisting with Coherent Light}
\author{M.Takeuchi}
 \affiliation{Department of Physics, Graduate School of Science, Kyoto University, Kyoto 606-8502, Japan}
\author{S.Ichihara}
 \affiliation{Department of Physics, Graduate School of Science, Kyoto University, Kyoto 606-8502, Japan}
\author{T.Takano}
 \affiliation{Department of Physics, Graduate School of Science, Kyoto University, Kyoto 606-8502, Japan}
\author{M.Kumakura}
 \affiliation{Department of Physics, Graduate School of Science, Kyoto University, Kyoto 606-8502, Japan}
 \affiliation{CREST, JST, 4-1-8 Honcho Kawaguchi, Saitama 332-0012, Japan}
\author{T.Yabuzaki}
 \affiliation{Department of Physics, Graduate School of Science, Kyoto University, Kyoto 606-8502, Japan}
\author{Y.Takahashi}
 \affiliation{Department of Physics, Graduate School of Science, Kyoto University, Kyoto 606-8502, Japan}
 \affiliation{CREST, JST, 4-1-8 Honcho Kawaguchi, Saitama 332-0012, Japan}
\date{\today}

\begin{abstract}
We propose a new method of spin squeezing of atomic spin,
 based on the interactions between atoms and off-resonant light
 which are known as paramagnetic Faraday rotation and fictitious
 magnetic field of light.
Since the projection process, squeezed light,
 or special interactions among the atoms are not required
 in this method, it can be widely applied to many systems.
The attainable range of the squeezing parameter is
 $\zeta\gtrsim S^{-2/5}$, where $S$ is the total spin,
 which is limited by additional fluctuations imposed by coherent light
 and the spherical nature of the spin distribution.
\end{abstract}

\maketitle
Squeezed spin state (SSS) is one of the non-classical
 states in collective spin system.
In SSS, the quantum uncertainty of the spins along an axis
 orthogonal to the mean spin vector
 $\langle\Delta\widetilde{S}_\perp^2\rangle$
 is suppressed below the standard quantum limit (SQL) such as 
 $\langle\Delta\widetilde{S}_\perp^2\rangle<|\langle\widetilde{\mathbf{S}}\rangle|/2$,
 where $\langle\widetilde{\mathbf{S}}\rangle$ is the mean spin vector,
 due to an entanglement formation among the individual spins.
The degree of the squeezing is usually evaluated
 by the squeezing parameter
 $\zeta\equiv2\langle\Delta\widetilde{S}_{\perp}^2\rangle/|\langle\widetilde{\mathbf{S}}\rangle|$,
 in terms of the variance to average ratio\cite{kitagawa}.

For the last several years, SSS
 has been extensively interested in not only
 for precision measurement of a spin component
 \cite{edm,itano,wineland},
 but also
 for the application to the quantum infomation\cite{lukin,polzik,andre}.
There have been many proposals and experiments
 to realize the spin squeezing of atoms.
 They can be put into three categories as follows :
 (i) Quantum non-demolition (QND) measurement of spin via
 paramagnetic Faraday rotation and
 spin squeezing by quantum projection
 \cite{polzik,mandel,takahashi,kuzmich,bigelow,feedback,mabuchi}.
 The QND measurement has been already performed by some groups
 for the electronic ground states of atom \cite{polzik,bigelow,mabuchi},
 and the squeezed parameter has reached about
 $\zeta\sim 0.7$ for $S\sim 4\times 10^7 $\cite{bigelow}, and 
 $\zeta\sim 0.1$ for $S\sim 10^{11}$\cite{mabuchi}.
 Since the projection causes the squeezing in this method,
 the degree of the squeezing will be finally determined
 by the performance of the detector.
 (ii) Quantum-state transfer from squeezed light to spin
 \cite{hald,absorb,lukin,andre,fleisch1,fleisch2}.
 One type is based on the complete absorption of squeezed
 vacuum, and has been experimentally demonstrated for
 the electronic exited states of atom
 ($\zeta\sim 0.97$ for $S\sim 5\times 10^7$) \cite{hald}.
 Another type is based on the stimulated Raman adiabatic passage
 \cite{lukin,andre,fleisch1,fleisch2}.
 Since the squeezed light is the source of spin squeezing in these
 methods, the degree of the squeezing will be finally determined
 by the quality of the squeezed light.
 (iii) Special systems to induce nonlinear interactions
 among the individual spins
 such as Bose-Einstein condensates\cite{bec-pdc,bec-oneaxis,beam},
 cold atoms in optical lattice\cite{lattice},
 atoms in optical cavity
 \cite{aps-search13,aps-search3,aps-search20}.
 They are not easy to prepare and difficult to operate after squeezed.

In this paper, we propose a new method to realize the spin squeezing,
 which can not be put into any of the three categories.
Our method does not rely on the projection by the measurement,
 use of squeezed light, and the specialities of the systems.
Instead, the new method only requires a coherent light pulse and a few
 linear optics, so it can be widely applied to many systems.
It should be noted that a recent electronic archive by 
 K.Hammerer \textit{et al.}\cite{hammerer} includes another proposal of
 an unconditional spin squeezing with coherent light.

Our method is based on the interaction
 between atoms and off-resonant light,
 whose interaction Hamiltomian takes a form \cite{takahashi}
\begin{align}
 H=\alpha J_z S_z,\label{Hgeneral}
\end{align}
 where $\alpha$ is a real constant,
 and the $z$-axis is set parallel to the wave vector of the light.
 $\mathbf{S}$ is the summation over the individual spin,
 which obeys the usual commutation relation of angular momenta
 $[\mathbf{S},\mathbf{S}]=i\mathbf{S}$.
$\mathbf{J}$ is quantum-mechanical Stokes vector of light,
 which also obeys the usual commutation relation of angular momenta
 $[\mathbf{J},\mathbf{J}]=i\mathbf{J}$.
For a light pulse with the duration $T$ propagating in free space,
 $\mathbf{J}$ can be written as 
$J_x \equiv \frac{1}{2} \int_0^T(a_\mathrm{+}^\dagger a_\mathrm{-}+a_\mathrm{-}^\dagger a_\mathrm{+})dt$,
$J_y \equiv \frac{1}{2i} 
\int_0^T(a_\mathrm{+}^\dagger a_\mathrm{-}-a_\mathrm{-}^\dagger a_\mathrm{+})dt$,
$J_z \equiv \frac{1}{2}
 \int_0^T(a_\mathrm{+}^\dagger a_\mathrm{+}-a_\mathrm{-}^\dagger a_\mathrm{-})dt$,
 where  $a_\pm$ is the annihilation operators of $\sigma_\pm$
 circular polarization mode, respectively \cite{duan}.
The interaction of Eq.(\ref{Hgeneral}) represents the addition of
 the phase difference for $\sigma_\pm$ light, which causes
 the rotation of the polarization plane for linear polarization
 at the angular frequency $\alpha S_z/2$,
 known as paramagnetic Faraday rotation.
 It also represents
 the spin rotation around the $z$-axis at the angular frequency 
 $\alpha J_z$, known as fictitious magnetic field of light\cite{fictitious}.
If we are able to apply a light pulse
 whose $J_z$ is proportional to $S_z$ as a fitctitious magnetic field,
 the collective spin will nonlinearly rotate
 at angular frequencies proportional to $S_z$, 
 whose evolution will be similar to one-axis twisting \cite{kitagawa}.
This is the basic idea of our proposal.

\begin{figure}[htbp]
 \includegraphics{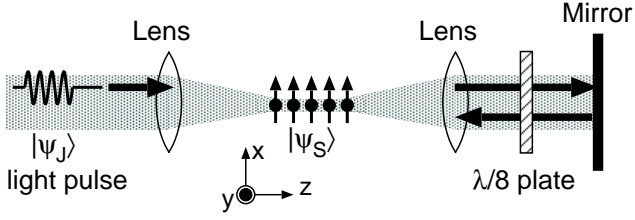}
 \caption{
 System of our proposal.
 A linearly polarized light pulse passes through
 an atomic ensemble and the polarization plane is rotated.
 The rotation angle is proporitonal to $S_z$ and 
 converted to the circular polarization components
 after passing through the $\lambda/8$ plate twice.
When the pulse passes through the atomic ensemble again,
 the pulse induce a nonlinear rotation
 to the atomic ensemble around the $z$-axis
 as a fictitious magnetic field.
 See text for details.
 }
 \label{setup.eps}
\end{figure}

To design such an interaction,
 we propose a system illustrated in Fig.\ref{setup.eps}.
Initially a light pulse $|\psi_J\rangle$ is linearly polarized
 along the $x$-axis and contains $2J(\gg 1)$ photons as an average.
Atoms $|\psi_S\rangle$ are spin-polarized
 along the $x$-axis and contains total spin $S$.
The light is weakly focused to match the atomic ensemble \cite{duan}.
The averages of the Stokes components is then $\langle J_x\rangle=J$,
 $\langle J_y\rangle=\langle J_z\rangle=0$, and 
 the averages of the collective spin vector is $\langle S_x\rangle=S$,
 $\langle S_y\rangle=\langle S_z\rangle=0$.
Since the light pulse is a strong coherent state,
 we can approximate the commutation relation as $[J_y,J_z]=iJ$ \cite{duan}.
Firstly, a light pulse passes through the atoms 
 and the polarization plane is then rotated.
We call it ``the first interaction'',
 whose interaction time is labelled as $t_1$.
The Stokes vector becomes
 $\mathbf{J}^\mathrm{(FI)}=e^{it_1H}\mathbf{J}e^{-it_1H}$,
 whose $y$ component is approximately written as
 $J_y^\mathrm{(FI)}\simeq J_y+\alpha t_1JS_z$,
 for $\alpha t_1 S_z \ll 1$.
Since the average of the $J_y^\mathrm{(FI)}$ becomes
 $\langle\psi_J|J_y^\mathrm{(FI)}|\psi_J\rangle=\alpha t_1JS_z$,
we can say that  the information of $S_z$ is copied 
 and held on $J_y^\mathrm{(FI)}$ as a Faraday rotation angle.
We note that $S_z$ is conserved
 because the interaction of Eq.(\ref{Hgeneral}) satisfies
 the back-action evasion (BAE) condition of $[S_z,H]=0$.
Secondly, the pulse passes through twice the $\lambda/8$ wave plate
 by the totally retroreflecting mirror.
As a result, $\lambda/4$ phase difference is induced between the two
 orthogonal  modes of linear polarization.
We call it ``the local operation'' for the light.
The Stokes vector becomes
 $\mathbf{J}^\mathrm{(LO)}=e^{i(\pi/2)J_x}\mathbf{J}^\mathrm{(FI)}e^{-i(\pi/2)J_x}$, whose $z$ component is
 $J_z^\mathrm{(LO)}=J_y^\mathrm{(FI)}$.
We can say that the information of $S_z$ is shifted
 from $J_y^\mathrm{(FI)}$ to $J_z^\mathrm{(LO)}$,
 converting the angle of the polarization plane to
 the photon number difference of the $\sigma_\pm$ modes.
Thus, the required light is achieved
 whose $J_z$ is approximately proportional to $S_z$.
Finally, the pulse passes through the atomic ensemble again.
We call it ``the second interaction'',
 whose interaction time is labelled as $t_2$.
The interaction Hamiltonian of the second interacton can be roughly written as
 $H^\mathrm{(SI)}\sim \alpha J_z^\mathrm{(LO)}S_z\propto S_z^2$,
 which takes a form similar to the one-axis twisting Hamiltonian
 $\chi S_z^2$
 \cite{kitagawa}.
Thus, we can expect that
 the spin state becomes SSS after the second interaction.

Next, we derive the density operator
 of the spin after the second interaction
 to calculate the properties of the spin state obtained by this method.
The initial density operator of the whole system can be written as
$\rho_{SJ}\equiv\rho_S\otimes|\psi_J\rangle\langle\psi_J|$,
 where $\rho_S=|\psi_S\rangle\langle\psi_S|$.
After the second interaction,
 it becomes
 $\widetilde{\rho}_{SJ}\equiv U\rho_{SJ} U^\dagger$
 where $U= e^{-it_2H}e^{-i(\pi/2)J_x}e^{-it_1H}$.
The reduced density operator representing the spin state
 after the second interaction $\widetilde{\rho}_S$ can be written as 
 $\widetilde{\rho}_S=\mathrm{Tr}_J(\widetilde{\rho}_{SJ})$,
 where $\mathrm{Tr}_J$ is the partial trace for the light.
For convenience, we consider the set of eigenstates
 for $\mathbf{S}^2$ and $S_z$, say $|S,M\rangle$,
 where $\mathbf{S}^2|S,M\rangle=S(S+1)|S,M\rangle$
 and $S_z|S,M\rangle=M|S,M\rangle$.
The matrix elements take a form
\begin{align}
 \langle S,M|\widetilde{\rho}_S|S,M'\rangle &=
 \sigma_{MM'}\langle S,M|\rho_S|S,M'\rangle, \label{recurrence}\\
 \sigma_{MM'}
 &= e^{-\mu'(M-M')^2/2}e^{-i\mu(M^2-M'^2)/2}, \label{sigma-ana}
\end{align}
 where we have set $\mu\equiv(\alpha t_1)(\alpha t_2)J$ and 
 $\mu'\equiv ((\alpha t_1)^2+(\alpha t_2)^2)J/2$.
If $t_1=t_2$ then $\mu=\mu'$.
Since the atoms $|\psi_S\rangle$ are polarized along the $x$-axis,
 the matrix elements of $\rho_S$ can be written as
\begin{align}
 \langle S,M|\rho_S|S,M'\rangle
 =\frac{1}{2^{2S}}\binom{2S}{S+M}^{1/2}\binom{2S}{S+M'}^{1/2} \label{rho_S}.
\end{align}
In the following discussions,
 we use the experessions of  Eq.(\ref{sigma-ana}) and Eq.(\ref{rho_S}). 
We note that
 the ideal one-axis twisted state \cite{kitagawa}
 corresponds to the case of $\mu'=0$.

\begin{figure}[htbp]
 \includegraphics{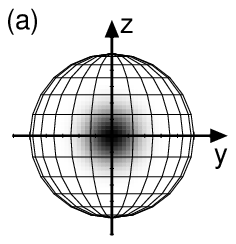}
 \includegraphics{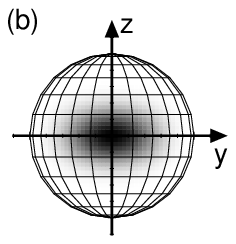}
 \includegraphics{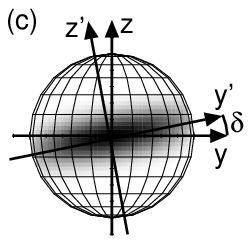}
 \caption{State evolutions 
 expressed as the quasiprobability distribution
 for $S=20$. 
 The value of QPD for $(\theta,\phi)$ direction
 is represented by the gray scale on the unit sphere,
 which is normalized by the maximum value.
 (a) The initial spin state.
 (b) The spin state after the first interaction,
 where we have set $(\alpha t_1)^2J/2=0.1$ and $t_2=0$,
 in other words, $\mu=0$ and $\mu'=0.1$.
 (c) The spin state after the second interaction,
 where we have set $(\alpha t_1)^2J/2=(\alpha t_2)^2J/2=0.1$.
 in other words, $\mu=\mu'=0.2$.
 The spin squeezing is realized along the $z'$-axis.
 }
 \label{QPD}
\end{figure}

To know how uncertainties evolve,
 we calculate the quasiprobability distributions (QPD),
 which is defined as $ Q(\theta,\phi)
 =\langle\theta,\phi|\widetilde{\rho}_S|\theta,\phi\rangle$,
 where $|\theta,\phi\rangle\equiv e^{-i\phi S_z}e^{-i\theta S_y}|S,S\rangle$
 is a spin state polarized along the direction
 whose polar and azimuth angles are $\theta$ and $\phi$,
 respectively \cite{kitagawa}.
The results of the calculations in the case of $S=20$ are shown
 in Fig.\ref{QPD} for the initial spin state (a),
 the spin state after the first interaction (b),
 and that after the second (c).
The initial spin state is isotropically distributed along
 the $x$-direction as is shown in Fig.\ref{QPD}(a).
After the first interaction, 
 the distribution is a little
 broadened along the $y$ direction, as Fig.\ref{QPD}(b) indicates.
This is explained by additional fluctuation imposed by coherent light.
In fact, the $y$ components after the first interaction
 is approximately written as
 $S_y^\mathrm{(FI)}=e^{it_1H}S_ye^{-it_1H}\simeq S_y+\alpha t_1 J_z S_x$
 for $\alpha t_1 J_z \ll 1$.
Since $S_z$ is the BAE variable, the distribution along the $z$ direction
 does not change at all.
After the second interaction, the distribution looks twisted
 around the $z$-axis and squeezed along the $z'$-axis,
 as Fig.\ref{QPD}(c) indicates.
Although not clear from the figure, 
 the distribution is also broadened along the $y$-axis
 as in the case after the first interaction.
In fact, the $y$ component after the second interaction
 is roughly written as $\widetilde{S}_y\sim e^{i
 t_2H^\mathrm{(SI)}}S_y^\mathrm{(FI)}e^{-it_2H^\mathrm{(SI)}}
 \sim S_y+\mu S_zS_x+(\alpha t_1J_z+\alpha t_2J_y)S_x$.
By these additional fluctuations imposed by coherent light,
 the spin state after the second interaction
 is different from the ideal one-axis twisted state \cite{kitagawa}.
The additional fluctuations
 would be reduced by use of a polarization
 squeezed light pulse whose squeezed component is
 $t_1J_z+t_2J_y$, approaching the ideal one-axis twisting
 interaction of $\mu'\to 0$.
We mention that the additional fluctuations by light in the method of
 Ref.\cite{hammerer} are imposed both on the  $y$ and $z$ 
 components, while the $z$ component is squeezed.
Therefore, the squeezing parameter does not become small in that scheme.

From Eq.(\ref{sigma-ana}) and Eq.(\ref{rho_S}), 
we can derive the averages and variances of the spin components.
The averages can be calculated as
$\langle\widetilde{S}_x\rangle=Se^{-\mu'/2}\cos^{2S-1}(\mu/2)$ and
$\langle\widetilde{S}_y\rangle=\langle\widetilde{S}_z\rangle=0$,
 where $\widetilde{\mathbf{S}}$ represents the spin operator after the
 second interaction.
 They indicates that the orientation of the mean spin vector remains
 $(\pi/2,0)$ direction or the $x$-axis, as is shown in Fig.\ref{QPD}(c).
To characterize a elliptical distribution around the $x$-axis,
 we define the minor and major axis,
 say $z'$ and $y'$, respectively,
 as is shown in Fig.\ref{QPD}(c),
 so that the variances of whose components
 $\langle\Delta\widetilde{S}_{z'}^2\rangle $
 and $\langle\Delta\widetilde{S}_{y'}^2\rangle $
 give the minimum and maximum on the $y-z$ plane, respectively.
The variances can be calculated as
$\langle\Delta\widetilde{S}_x^2\rangle=S^2-\langle\widetilde{S}_x\rangle^2-S(S-1/2)A/2$
and
\begin{align}
 \langle\Delta\widetilde{S}_{\genfrac{}{}{0pt}{}{y'}{z'}}^2\rangle
 &=\frac{S}{2}+\frac{S}{2}\frac{S-1/2}{2}
 \left(A\pm\sqrt{A^2+B^2}\right),
\end{align}
 where we have set
 $A=1-e^{-2\mu'}\cos^{2S-2}\mu$, and
 $B=4e^{-\mu'/2}\sin(\mu/2)\cos^{2S-2}(\mu/2)$.
We can also calculate $\delta$,
 which is an angle between the directions of the $z'$- and $z$-axes,
 or the $y'$- and $y$-axes, as is shown in Fig.\ref{QPD}(c),
 and obtain $\delta=\arctan(B/A)/2$.
For $S\gg 1$ and $S^{-1}\ll \mu\sim\mu'\ll S^{-1/2}$, 
 we find the approximate value of the variance of the $z'$ component
\begin{align}
 \langle\Delta\widetilde{S}_{z'}^2\rangle
 &\simeq\frac{S}{2}
 \left(\frac{\gamma'}{\gamma^2+\gamma'}+\frac{2}{3}\beta^2\right), \label{Sz'}
\end{align}
where we have set $\gamma=S\mu/2$, $\gamma'=S\mu'/2$ and
 $\beta=S\mu^2/4$.
Also we find $\langle\widetilde{S}_x\rangle\simeq S(1-\beta)$.

\begin{figure}[htbp]
 \includegraphics{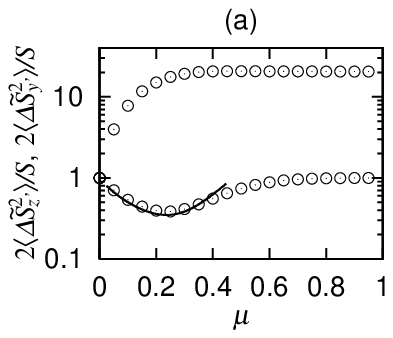}
 \includegraphics{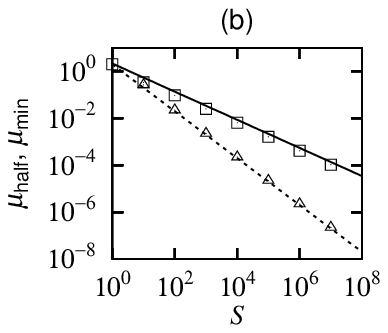}
 \caption{
 (a)Variances of the $z'$ (circle below 1)
 and the $y'$ (circle above 1) components for $S=20$ as a function of $\mu$
 ($\mu_\mathrm{half}=0.117$, $\mu_\mathrm{min}=0.236$).
 They are normalized
 as $2\langle\Delta\widetilde{S}_{z'}^2\rangle/S$
 and $2\langle\Delta\widetilde{S}_{y'}\rangle/S$,
 respectively.
We also plot the approximate
 value of $2\langle\Delta S_{z'}^2\rangle/S$ (solid line).
 (b)Value of  $\mu_\mathrm{half}$ (triangle) and
  $\mu_\mathrm{min}$ (square) as a function of $S$,
 which are the required values of $\mu$
 to obtain the half variance of the SQL and
 the attainable minimum variance, respectively.
 We also plot the approximate solutions of
 $\mu_\mathrm{half}$ (dashed line) and $\mu_\mathrm{min}$ (solid line).
 To obtain both (a) and (b), we have assumed $\mu=\mu'$.
 }
 \label{variance}
\end{figure}

To examine the dependence on the interaction strength
 $\alpha t_1,\alpha t_2$ and the input photon number $2J$,
 we plot the variances of the $y'$ components
 and the $z'$ components 
 as a function of $\mu(=\mu')$ in Fig.\ref{variance}(a).
We also plot the approximate value for the  $z'$ components
 written as Eq.(\ref{Sz'}).
It is clearly known that 
 the variance of the $z'$ component
 is reduced for small $\mu$, minimized at an optimal value of $\mu$,
 and becomes large for large $\mu$.
It means that
 too strong interaction or too large photon number deteriorates
 the squeezing.
This is explained by the spherical nature of the spin distribution,
 and in fact, the variance of the $y'$ component
 is almost saturated at the largest value of $S^2/2$
 for large $\mu$, which was entirely ignored in the analysis 
in Ref.\cite{hammerer}.
As a typical value of $\mu$,
 we introduce $\mu_\mathrm{half}$ as the value of $\mu$
 to attain $\langle\Delta\widetilde{S}_{z'}^2\rangle=S/4$,
 the half variance of the SQL.
We also introduce $\mu_\mathrm{min}$ as the value
 to attain the minimum of the $\langle\Delta\widetilde{S}_{z'}^2\rangle$.
We plot the numerical solutions of $\mu_\mathrm{half}$ and
 $\mu_\mathrm{min}$ in Fig.\ref{variance} (b) for the case of $\mu=\mu'$.
One can see that both $\mu_\mathrm{half}$ and $\mu_\mathrm{min}$ become 
 small as $S$ increases but they obey different power laws.
From Eq.(\ref{Sz'}), we find
$\mu_\mathrm{half}\simeq 2S^{-1}$ and
$\mu_\mathrm{min}\simeq 2(3/2)^{1/5} S^{-3/5}$.
We show these approximate solutions in Fig.\ref{variance} (b),
 which are in good agreement with the numerical ones.
We also find that the squeezing parameter at $\mu=\mu'=\mu_\mathrm{min}$
 becomes $\zeta_\mathrm{min}\simeq (2/3)^{1/5}S^{-2/5}$.
We note that it is slightly worse than
 $(1/3)^{1/3}S^{-2/3}$, which is the squeezing parameter
 for the ideal one-axis twisting, 
 due to the additional fluctuations imposed by coherent light,
 as is mentioned above.

Finally, we discuss the feasibility of our method.
In the following,
 we consider the case that the shape of the light pulse is a square wave
 with its peak power $P$ and pulse duration $T$.
As in Ref.\cite{duan}, 
 we assume $\Delta \gg \Gamma$, 
 $\Omega \ll \Delta$, and $rT \ll (S\mu)^{-1}\ll 1$, 
 where
 $\Delta$ represents the detuning from the resonance frequency,
 $\Gamma$ the full natural linewidth at half maximum of the transition,
 $\Omega$ the Rabi frequency,
 and $r$ the photon scattering rate \cite{suter}.
After some calculations, we obtain
$\mu=\mu'=rT\sigma_0/(2\pi w^2)$,
 where $w$ represents the beam waist and
 $\sigma_0$ the photon-absorption cross section of an
 atom, which can be written as $\sigma_0=3\lambda_0^2/(2\pi)$ with the
 resonance wavelength $\lambda_0$.
We note that $\Omega$, $r$ and $\mu$ are exactly the same
 as $2g\sqrt{2N_p/(cT)}$, $4\varepsilon_a/T$, and $\kappa^2/N_a$
 in Ref.\cite{duan}, respectively.
The condition to obtain $\mu\ge\mu_\mathrm{half}$,
 or $\zeta\le 1/2$, can be rewritten as $d_0\gtrsim 8(rT)^{-1}$,
 where $d_0= 2S\sigma_0/(\pi w^2)$ is the optical depth.
We also note that 
 this condition is the same as $\kappa\gtrsim\sqrt{2}$, similar
 to that of QND measurement \cite{duan}.
Such a condition has been satisfied in several systems,
 such as atoms in a cell \cite{polzik},
 laser cooled and trapped atom, and so on.
The feasibility of our scheme also comes from the simple experimental
setup depicted in Fig.\ref{setup.eps}, which is also the great advantage
over another scheme in Ref.\cite{hammerer}.
This suggests efficient squeezing can be realized
 by the current technologies.

As one ideal example, we consider a ytterbium atoms (${}^{171}$Yb)
 in optical trap\cite{honda,takasu}, which contains $S= 4\times 10^{6}$.
The atom collision and the precession due to the stray magnetic field,
 which causes the transverse relaxation, are well surpressed
 because it is ultracold fermion and
 has only nuclear spin 1/2 whose gyromagnetic ratio is about three orders
 smaller than paramagnetic atoms like alkali metal.
From the parameters given in Ref.\cite{takasu},
 $w= 3\mathrm{\mu m}$ \cite{takasu}, 
 $\lambda_0=399\mathrm{nm}$ and
 $\Gamma=2\pi\times29\mathrm{MHz}$,
 the light pulse of $\mu=5.4\times 10^{-6}$, for example,
 is obtained by setting
 $\Delta=2\pi\times 24$GHz,
 $P=17\mathrm{n W}$,
 $T=0.24\mathrm{m s}$,
 which satisfies the assumptions
 $\Delta \gg\Gamma$,
 $\Omega=2\pi\times 21\mathrm{MHz}\ll\Delta$,
 $rT=4.0\times 10^{-3}\ll (S\mu)^{-1}=4.6\times 10^{-2}\ll 1$.
 and $J=4.0\times 10^6\gg 1$.
In this case the squeezing parameter becomes $\zeta=0.08$.
We note that the decay constant of the atom number
 in optical trap is about $4$s \cite{ybbec},
 which is so longer than the the pulse duration $T$ that
 we can treat the total spin $S$ as a constant.
We also note that the length of the atom distribution shuould be adjusted to
 $L\sim 70\mathrm{\mu m}$ to satisfy the condition
 of $\pi w^2/(\lambda_0 L)\sim 1$\cite{duan}, which is easy for atoms in
optical trap of crossed configuration.

To avoid the interference between 
 the three steps of the first interaction, the local operation and
 the second interaction within one pulse,
 we can use a pluse train, each duration of which is so short
 that the three steps are separable and the repetition rate is so slow that
 the pulse number travelling in the path is at most one.
Since the each matrix element $\langle S,M|\rho_S|S,M'\rangle$
 would evolve like a geometric progression
 whose common ratio is $\sigma_{MM'}$ for the every pulse
 as Eq.(\ref{recurrence}) indicates,
 and $\sigma_{MM'}$ is the power of
 the mean photon number of the each pulse
 as Eq.(\ref{sigma-ana}) indicates,
 we can say that the same SSS would be obtained
 as long as the total mean photon number passed through the atomic
 ensemble are equal.

\end{document}